# IMPROVING PHISHING DETECTION VIA PSYCHOLOGICAL TRAIT SCORING

Sadat Shahriar, Arjun Mukherjee, Omprakash Gnawali
*University of Houston*
*4800 Calhoun Rd, Houston, TX 77004, USA*

**ABSTRACT**

Phishing emails exhibit some unique psychological traits which are not present in legitimate emails. From empirical analysis and previous research, we find three psychological traits most dominant in Phishing emails – **A Sense of Urgency**, **Inducing Fear by Threatening**, and **Enticement with Desire**. We manually label 10% of all phishing emails in our training dataset for these three traits. We leverage that knowledge by training BERT, Sentence-BERT (SBERT), and Character-level-CNN models and capturing the nuances via the last layers that form the Phishing **Psychological Trait (PPT)** scores. For the phishing email detection task, we use the pretrained BERT and SBERT model, and concatenate the PPT scores to feed into a fully-connected neural network model. Our results show that the addition of PPT scores improves the model performance significantly, thus indicating the effectiveness of PPT scores in capturing the psychological nuances. Furthermore, to mitigate the effect of the imbalanced training dataset, we use the GPT-2 model to generate phishing emails (Radford et al., 2019. Our best model outperforms the current State-of-the-Art (SOTA) model's F1-score by 4.54%. Additionally, our analysis of individual PPTs suggests that Fear provides the strongest cue in detecting phishing emails.

## 1. INTRODUCTION

Phishing is a technique used in electronic messaging to deceive the reader, where the phisher camouflages the message with a legitimate facade to access sensitive information or monetary gain (Vishwanath et al., 2011; Bose and Leung 2009). As the phishers prey on the vulnerability of the users, they often persuade people to take on some actions which may lead to undesirable consequences. Phishing attacks increased significantly in recent years and although researchers exploited several Natural Language Processing and Machine Learning techniques to detect phishing emails, the phishers evolved over time, making it harder to detect phishing emails (Almomani et al., 2013; Khonji, Iraqi, and Jones, 2013; FBI, 2021). Hence, phishing email detection systems must be smart enough to cope with the evolving nature of phishing techniques. In this work, we propose that all phishing emails exhibit some unique psychological traits, and detection of these traits can play a significant role in improving phishing vs. legitimate email classification.

Researchers analyzed the phishing attack from psychological perspectives, such as how persuasion is conducted (Akbar, 2014; Cialdini, 2001), the human factors in phishing attack (Stajano and Wilson, 2011; Jakobsson, 2007), and psychological mechanism in the effectiveness of phishing attacks (Luo et al., 2013). Research suggests that phishing messages often exhibit psychological cues, which can be crucial for their successful detection (Jones et al., 2019; Jakobsson, 2007). However, the current research is not adequate to quantify psychological traits expressed through the body of text. Consequently, how these traits play into detecting phishing emails is still an unexplored area of research. Nevertheless, for a smart detection of phishing emails, it is of immense importance to incorporate psychological attributes of the email's text, along with the linguistic model. We define *Phishing Psychological Traits (PPT)* as the psychological attributes evident in phishing emails. We claim that three major psychological attributes are evident in the phishing emails–based on whether the email sounds rushed (*a Sense of Urgency*), if the email induces fear *(Inducing Fear by*

*Threatening*), and if there is an enticement through that email (*Enticement with Desire*). These traits can appear standalone or with a combination of any two and even three.

In this research, we capture the psychological traits by modeling a BERT, Sentence-BERT (SBERT) and Char-CNN network (Devlin et al., 2019; Reimers and Gurevych, 2019; Zhang, Zhao, and LeCun, 2015). We use these models to compute the softmax probability score (PPT score) for every phishing and legitimate emails. Next, we use pretrained BERT and SBERT model to find the feature-embedding (768-D) from text and concatenate the PPT scores with these embeddings. The concatenated features are fed to a fully-connected neural network to predict the email being phishing or legitimate. Our best performing model achieves the F1-score of 88.04%, which outperforms the current SOTA by 4.54%. We also observe a significant improvement of F1- score by up to 2.62% for the PPT-based model over the PPT-less model. The key to the consistent performance improvement is the PPT scores which provide reliable and unique cues by capturing the subtlety of psychological aspect expressed in the emails.

The novelty of this research is that our work is the first one to quantify the underlying psychological cues and leverage them for phishing email detection. We further investigate how the PPT scores help boosting the classification performance by providing t-SNE-based visualization (van der Maaten and Hinton, 2008). Furthermore, we analyze the contribution of individual PPT score and effectiveness of PPT scores in low-training-data situations. Our research provides important insights into the unique psychological attributes of phishing emails, which can create a new research direction in the phishing email detection paradigm.

## 2. RELATED WORKS

Phishing emails have been adversely affecting the internet world since 1996 (Salloum et al., 2021). Different NLP techniques were used to extract semantic, syntactic and contextual features which played important role for phishing detection along with classical machine learning techniques (Cui et al., 2020; Verma and Hossain, 2013; Park and Taylor, 2015; Blanzieri and Bryl, 2009; Gansterer and Pölz, 2009; Feng et al., 2016). However, the evolving nature of phishing emails entails more sophisticated techniques as often they do not contain the malicious code or word choices of known attacks (Lee, Saxe, and Harang, 2020a). The transformer-based models and their variants proved to have superiority over the traditional deep learning models mostly due to their transfer learning capabilities and they were successfully used in many deceptive text detection tasks (Vaswani et al., 2017; Shahriar et al., 2021). Lee et al. proposed a BERT-based phishing email detection model where they pruned half of the transformer blocks to better capture the semantics (Lee, Saxe, and Harang, 2020b). However, none of these works consider the unique psychological aspect of phishing emails, that can provide useful features to detect them. Naidoo suggested the *urgency* to be a dominant psychological feature in phishing email, but an ML-based automatic detection strategy is not present in their work (Naidoo, 2015). We address these research gaps by leveraging the context embedding of the BERT and SBERT model and using the psychological traits to detect phishing emails.

## 3. METHODOLOGY

Figure 1 explains the whole process in brief. We start by selecting 10% of the phishing emails and manually label them as 1 or 0 based on the presence of each PPTs. Next, we train a BERT, SBERT and Char-CNN model for each PPTs and use the trained model to compute the PPT scores of all emails. We concatenate these PPT scores with the fine-tuned BERT model and pretrained SBERT model and feed them to a Deep Neural Net model to detect the phishing email.

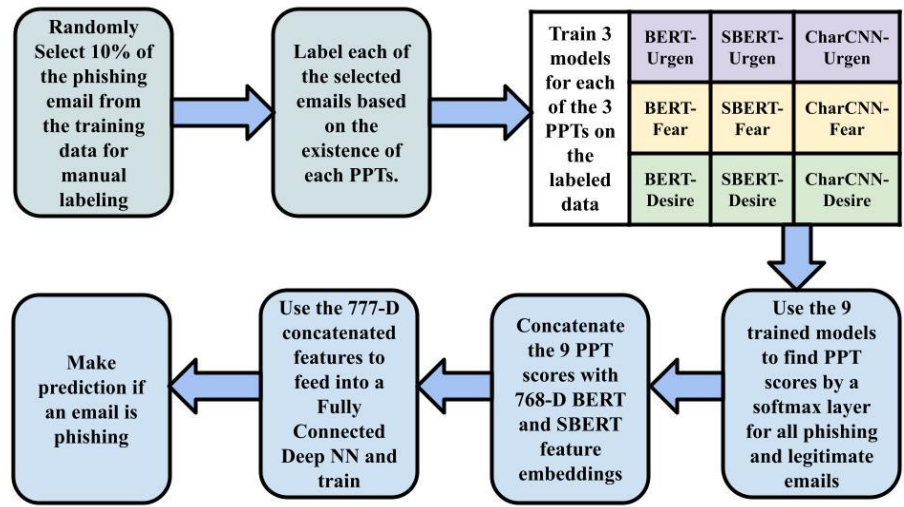

Figure 1: The complete flow diagram for phishing email detection using Phishing Psychological Traits (PPT)

## 3.1 Phishing Psychological Traits (PPT)

We claim that one or more persuasion strategies are implemented with the psychological traits expressed in the text. PPTs work as a broader umbrella and all phishing emails exhibit one or more of these PPTs.

### 3.1.1 A Sense of Urgency

One of the crucial characteristics of many phishing messages is the expression of urgency. Having a time constraint can induce stress in the readers' mind even if they are capable to do so within the stated time (Ordonez and Benson III, 1997). The phishing emails often urge the reader to take some action with promptness, thus reducing the time for the reader to reason or report (Aggarwal, Kumar, and Sudarsan, 2014). Additionally, it can induce impulsive behavior in the recipient that can lead to an error of judgment (Cui et al., 2020). Therefore, it is imperative to detect urgency from an email. We observe that urgency expressed in the phishing messages are direct and expressive, as the attackers want to maximize the possibility of a reader's response.

### 3.1.2 Inducing Fear by Threatening

Research shows that Fear by threatening is one of the most frequently exploited emotional trigger by the attackers (Sharma and Bashir, 2020, Halevi, Lewis, and Memon, 2013; Ferreira and Lenzini, 2015). The threat can be of many forms, for example, being locked out or blocked from one's account, losing access to information, getting hacked, stealing information, stealing currency, and individually targeted attack (Wang et al., 2012). Notably, Bitaab et al. stated that during the COVID-19 pandemic, the readers' fear is exploited by the attackers, which led to a high increase of phishing attacks (Bitaab et al., 2021). Hence it is evident that a direct or indirect threat can be a significant cue of phishing emails.

### 3.1.3 Enticement with Desire

Phishing email often lures by enticing the readers' personality trait of openness that can make them to be greedy and curious which can result in getting phished (Ding et al., 2015; Halevi, Lewis, and Memon, 2013). Phishing emails often contain a lucrative financial reward in exchange for clicking on some links, providing personal details or credit card information, and so on. Stajano and Wilson maintained that "Need and Greed" is one of the seven basic principles of scams, where people can be a victim of a lottery scam or a sexy swindler (Stajano and Wilson, 2011). Hence, the enticement with greed or curiosity can be an important signal of phishing emails.

## 3.2 Experimental Setup

To obtain the PPT scores of all emails, we train three models for each of the PPTs – BERT, SBERT and Char-CNN. We split all the manually-labeled-PPT emails in 80% for training and 20% for validation and repeat the experiments for three different splits. For the training of phishing email detection task, we use the BERT and SBERT model and apply the same train-validation split. We find the best set of hyperparameters by observing the performance on the validation set.

## 4. DATASET

The dataset we used was provided in the Anti-Phishing Pilot at ACM IWSPA 2018 (Verma, Zeng, and Faridi, 2019). Our test data size is 4300, from the first shared task, where emails are provided without the headers (*IWSPA_NH*). The IWSPA_NH training set has 5092 legitimate and 629 phishing emails. We also added 4082 legit, 503 phish emails from the IWSPA header-added dataset (*IWSPA_H*). Additionally, to examine how our trained model performs on other datasets, we curated a new small dataset called *UNIV_Phish*, containing 326 emails. We collected 163 phishing email from three different university websites: 72 emails from Stanford University, 68 from Lehigh University and 23 from University of Washington. We also added 163 emails from Enron "ham" emails. Notably, we made sure, none of these emails appeared in the IWSPA training set. The curated data can be found here: https://github.com/sadat1971/Phishing_Email

## 5. RESULT AND DISCUSSION

We hypothesize that when we add the PPTs along with the language model, the performance of phishing email detection improves. Therefore, we first need to find the PPT scores for all emails and then use these scores to detect phishing emails.

### 5.1 Detection of Phishing Psychological Traits

The randomly selected phishing emails are labeled by one of the authors as 1 or 0 for each PPTs. We find 82.54% emails are labeled as Urgent, where 71.42% emails are labeled as both Urgent and Fear. Only 4.76% emails exhibit all the three traits. We use BERT, SBERT, and Char-CNN to train on the manually labeled emails. Then, we make prediction on rest of the email using these models' last softmax layer and use the softmax output as PPT score.

Figure 2 depicts the distribution of BERT-based PPT scores in phishing vs legitimate emails. We observe that the PPT scores create distinguishable clouds for phishing and legitimate emails in Figure 2(a). Figure 2(b) shows the kernel density plot of individual PPTs, which represents the continuous probability density curve for these traits. For *Sense of Urgency* and *Fear by Threatening*, we observe a high density of Phishing emails on the right side of the curve which indicates the high probability of *urgency* and *fear* in the phishing emails. However, contrary to our intuition, phishing emails have a lower probability in the *Desire* score than the legitimate emails. One reason could be the lack of phishing emails in our dataset with this trait, which may have caused the failure to capture the *Desire* trait in the emails.

We explore how different words are related to the three PPTs. We observe "immediately", and "soon" are two most recurring adverbs in the *Urgent*-labeled emails. The *Desire*-labeled emails are mostly the offers of a free upgrade of some subscriptions. However, since the main goal of phishing emails is to persuade the user to take some action by clicking on a phishing link, we observe the words "link" and "click" frequently in all the phishing emails.

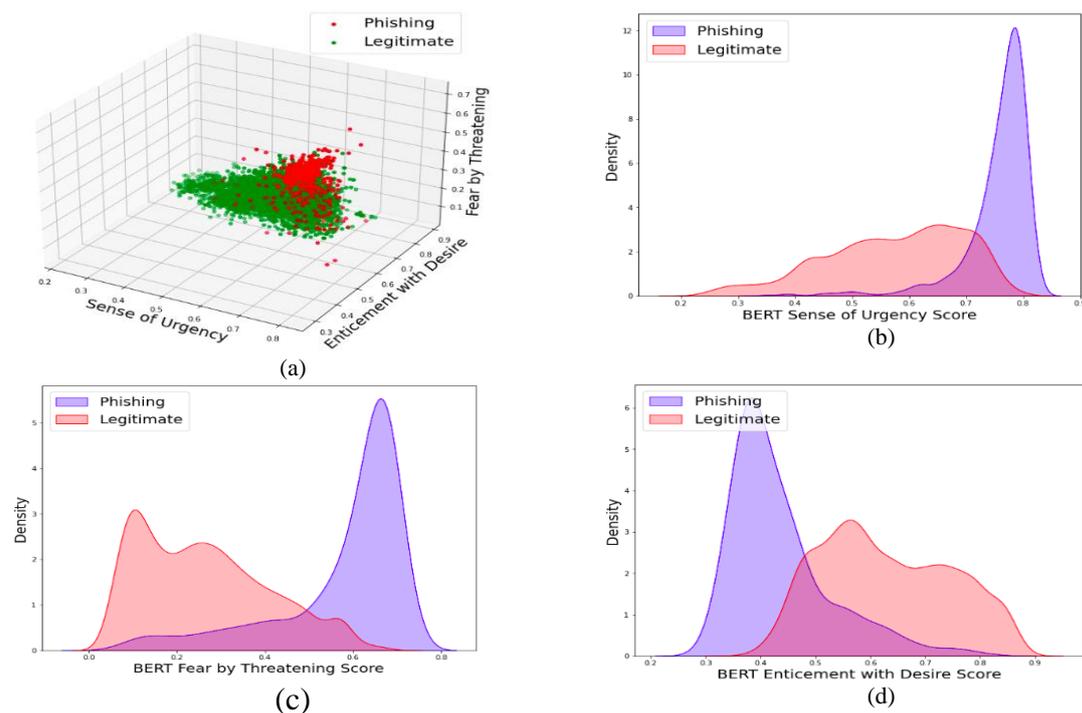

Figure 2: (a) 3-dimensional plot of PPT scores for BERT. (b) Kernel Density Estimation (KDE) plot of BERT-based PPT Score for Sense of Urgency (c) KDE plot Fear by Threatening (d) KDE plot for Enticement with Desire.

## 5.2 Phishing Email Detection

From Table 1, we observe that, for every case, the performance improves when PPT scores are added to the training process. While trained on IWSPA_NH training data for the BERT model, we found that adding the PPTs improves the accuracy by 0.70% and F1-score significantly by 2.62% (p-value=0.04). We find the same trend for SBERT model as well. Adding the psychological traits improves the accuracy by 1.31% and F1-score by 1.34%.

Next, we added the additional training data for IWSPA with header information (IWSPA_H). The additional training data helps improve the performance. However, adding the PPTs again improves the performance. For BERT model, the improvement is by 0.63% in accuracy and 2.19% (p-value=0.03) in F1-score. Similarly, for the SBERT model, F1-score has a significant improvement of 2.50% (P-value=0.02). It may be noted that the current SOTA F1-score for this test set is 83.5%. Adding the PPTs with IWSPA_NH and IWSPA_H training data, we outperform the current SOTA (85.16% for BERT and 83.88% for SBERT).

A major challenge in our task is the lack of training data in the phishing email category due to the corporations being reluctant and individuals being ashamed to share such sensitive data (Aassal et al., 2018). In order to balance the training dataset, we tried different approaches like SMOTE (Chawla et al., 2002), cost-sensitive learning methods (Thai-Nghe, Gantner, and Schmidt-Thieme, 2010), and the addition of GPT- 2-generated phishing emails (Radford et al., 2019). However, we did not find any performance improvement for the first two. For GPT-2, we observe the performance boost for BERT models by 1.02% in accuracy and 3.59% in F1-score. When we added the psychological trait features with GPT- 2-generated emails, it also improved the accuracy by 0.45%, and F1-score by 1.38% .

Table 1. Performance of the IWSPA test set and UNIV_Phish dataset while trained on different training data. We observe the best performance is found when we use IWSPA header-less, header-added data, GPT-2-generated phishing emails, and PPT features added with the BERT model

| Tested On | Training Data | BERT | | BERT + PPT | | SBERT | | SBERT + PPT | |
|---|---|---|---|---|---|---|---|---|---|
| | | Acc (%) | F1 (%) | Acc (%) | F1 (%) | Acc (%) | F1 (%) | Acc (%) | F1 (%) |
| IWSPA Test set | *IWSPA_NH* | 95.11 | 79.34 | 95.81 | 81.96 | 95.39 | 79.29 | 95.70 | 80.63 |
| | *IWSPA_NH + IWSPA_H* | 96.00 | 83.07 | 96.63 | 85.61 | 96.23 | 81.38 | 96.54 | 83.88 |
| | *IWSPA_NH+IWSPA_H+GPT2* | 97.02 | 86.66 | 97.47 | **88.04** | 94.32 | 77.19 | 95.04 | 79.41 |
| UNIV_ Phish | *IWSPA_NH* | 85.27 | 84.71 | 86.81 | 85.61 | 81.29 | 79.38 | 82.82 | 80.70 |
| | *IWSPA_NH + IWSPA_H* | 86.50 | 85.23 | 87.11 | 86.17 | 82.21 | 80.01 | 83.74 | 82.15 |
| | *IWSPA_NH+IWSPA_H+GPT2* | 87.42 | 86.98 | 88.03 | **87.77** | 80.06 | 76.63 | 82.82 | 80.55 |

However, contrary to the BERT model, the addition of GPT-2 generated data created a significant decline in the SBERT performance. The reason could be the poor coherence of some of the generated data. While for the BERT model, we use the output from the [*cls*] token, SBERT uses a pooling strategy, leading to poor sentence embedding of non-coherent texts. Additionally, as suggested in Reimers and Gurveych (Reimers and Gurevych, 2019), since SBERT cannot be used to update all the internal layers of BERT architecture, it may not be well suited for transfer learning.

We further test our model performance on UNIV_Phish dataset. From Table 1, we observe that added PPT improves the performance up to 4.02% in F1-score. We also observe the similar performance boost with added IWSPA_H set (up to 1.45%) and added GPT-2-generated email set (up to 1.75%). Hence, the performance of UNIV_Phish dataset further strengthens our model validity.

Next, we analyze the embedding representation of the emails using t-SNE plot (Maaten and Hinton, 2008). Figure 3 shows that the cloud of misclassified samples is mainly in the overlapped region of phishing and legitimate emails, which indicates the lack of better embedding representation of these emails. However, we observe that the distance between the center of phishing email and the legitimate email cloud increased by adding the PPTs by 9.56%, and 15.29% using Euclidean and Manhattan distance, respectively. Thus, the addition of psychological traits seems to improve the embedding representation.

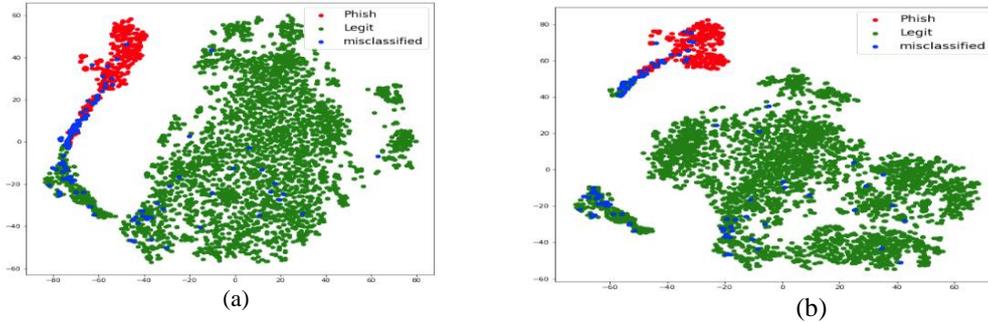

Figure 3: t-SNE representation of Phishing and Legitimate email with (a) BERT-based feature only (b) BERT-based + Phishing Psychological Trait features. The figure shows the misclassification zones in blue.

We further examine the effectiveness of the Phishing Psychological Traits model by ablation experiments (Meyes et al. 2019). From Figure 4 (a), we observe that the performance decreases the most, when we remove the *Urgency* trait (0.91% in F1-score, 1.01% in accuracy), followed by the *Fear* trait (0.88% in F1-score, 0.99% in accuracy). *Desire* had the least effect on performance (0.60% in F1-score and 0.71% in accuracy), which is consistent with our previous analysis.

Finally, we vary the training data proportion to examine the effect of PPT when we have a small amount of training data. Figure 4(b) shows that while with 100% training data, PPT scores improve the F1-score by 2.62%, with only 20% training data, the F1-score improvement is by 3.94% indicating the effectiveness of PPTs even with insufficient training data. Figure 4(b) also demonstrates the effect of adding PPTs individually. We observe that as a standalone PPT, Fear by Threatening has a better impact on performance than the others. Nevertheless, the three PPTs combined provide the best cue for detecting phishing emails.

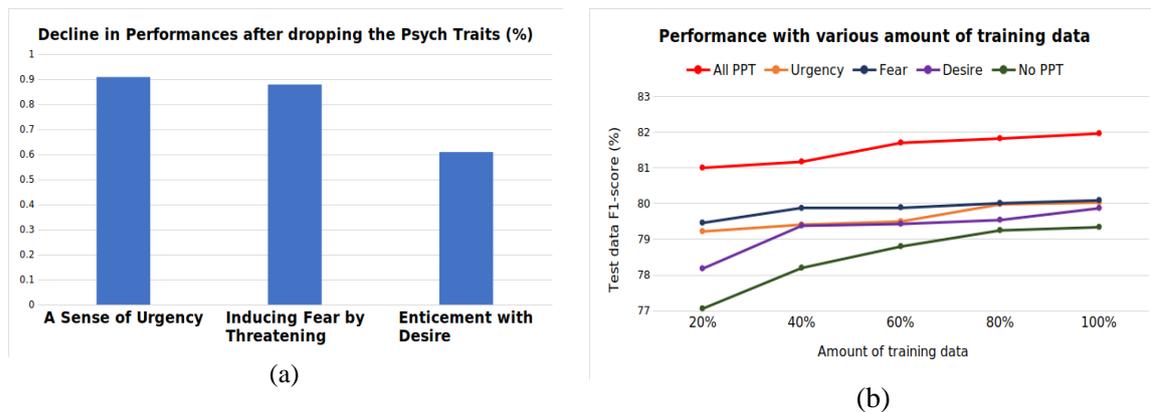

Figure 4: (a) Decline in performance after dropping the PPTs one at a time (b) Performance in test data at varying proportion of training data

## 6. CONCLUSION

Quantifying the psychological traits of an email can provide key signals which help improve the phishing email detection performance. In this paper, we define, analyze, and quantify the PPTs that can successfully capture the nuances of an email's intent and show promising results. Hence, our work may provide potential research direction to win the battle against evolving nature of phishing. However, we still have limitations and room for further improvement. First, we will obtain ground truth for the PPTs by labeling them with multiple human raters enabling us to measure the kappa statistics for testing inter-rater reliability, which in turn can provide a more accurate estimation of PPTs. Second, further research might be required to understand the flow of psychological traits in conversational turns to detect more organized phishing than single email-based phishing. Finally, an investigation of how the individual PPTs contribute to forming a phishing email can provide valuable insight that can be utilized for more efficient phishing email detection.

## ACKNOWLEDGMENTS

Research was supported in part by grants NSF 1838147, ARO W911NF-20-1-0254. The views and conclusions contained in this document are those of the authors and not of the sponsors. The U.S. Government is authorized to reproduce and distribute reprints for Government purposes notwithstanding any copyright notation herein.## REFERENCES